\documentclass[journal]{IEEEtran}
\input{ds.def}

\newcommand{\AQLNGS}{INFN Laboratori Nazionali del Gran Sasso, Assergi (AQ) 67100, Italy}
\newcommand{\AQGSSI}{Gran Sasso Science Institute, L'Aquila 67100, Italy}

\newcommand{\Houston}{Department of Physics, University of Houston, Houston, TX 77204, USA}

\newcommand{\Princeton}{Physics Department, Princeton University, Princeton, NJ 08544, USA}

\begin{document}
\title{Development of a novel single-channel, 24~cm$^2$, SiPM-based, cryogenic photodetector}
\author{
  	Marco~D'Incecco,
	Cristiano~Galbiati,
	Graham~K.~Giovanetti,
	George~Korga,
	Xinran Li,
	Andrea~Mandarano,
	Alessandro~Razeto,
	Davide~Sablone,
	and~Claudio~Savarese
	\thanks{Manuscript revisioned on \today.}
	\thanks{We acknowledge support from \NSF\ (US, Grant PHY-1314507 for Princeton University), the Istituto Nazionale di Fisica Nucleare (Italy) and Laboratori Nazionali del Gran Sasso (Italy) of INFN.}
	\thanks{Work at Princeton University was supported by Fermilab under Department of Energy contract DE-AC02-07CH11359.}
	\thanks{M.~D'Incecco is with \AQLNGS.}
	\thanks{C.~Galbiati, G.K.~Giovanetti and X.~Li are with \Princeton.}
	\thanks{G.~Korga is with \Houston\ and \AQLNGS.}
	\thanks{A.~Mandarano, and C.~Savarese are with \AQGSSI\ and \AQLNGS.}
	\thanks{D.~Sablone and A.~Razeto are with \AQLNGS\ and \Princeton.}
	\thanks{Corresponding Author: sarlabb7@lngs.infn.it}
}
\maketitle
\begin{abstract}
We report on the realization of a novel \SiPM-based, cryogenic photosensor with an active area of 24~cm$^2$ that operates as a single-channel analog detector. The device is capable of single photon counting with a signal to noise ratio better than 13, a dark rate lower than $10^-2$~cps/mm$^2$ and an overall photon detection efficiency significantly larger than traditional photomultiplier tubes. This development makes SiPM-based photosensors strong candidates for the next generation of dark matter and neutrino detectors, which will require multiple square meters of photosensitive area, low levels of intrinsic radioactivity and a limited number of detector channels.
\end{abstract}
\IEEEpeerreviewmaketitle
\section{Introduction}
\label{sec:intro}
The excellent single-photon resolution, photon detection efficiency, cryogenic performance and cost of silicon photomultipliers (\SiPMs) makes them appealing replacements for conventional photomultiplier tubes in next generation physics experiments. However, SiPMs typically have active areas of several tens of square millimeters, meaning a large number of devices is required to instrument the area of a single photomultiplier tube. This poses significant space and interconnection challenges in large experiments, where readout cables are long and front-end electronics must be placed in close proximity to the detector. An extra layer of complication arises in rare-event searches, such as those searching for neutrinoless double-beta decay or dark matter, where the background contributions from cables and electronics are often dominant. One way to alleviate this problem is to group \SiPMs\ together to reach a surface area equivalent to a photomultiplier tube and read them out as a single channel device. However, the high capacitance of \SiPMs, about \SI{50}{pF\per\square\mm}, makes it challenging to operate a large array of \SiPMs\ as a single channel device while maintaining adequate timing resolution and single photon resolution.

\subsection{Goals}
This article focuses on the realization of a cryogenic, single analog output, \SiPM-based photodetector with an active area comparable to a \SI{3}{\inch} photomultiplier tube (PMT) and better performance in terms of photon detection efficiency (PDE) and single photon resolution. 

The \SiPM\ layout and the front-end electronics have been designed to achieve a signal to noise ratio (SNR), defined here as the ratio between the amplitude of the single photo-electron signal and the standard deviation of the baseline noise, better than 10. At this noise level, the rate of baseline fluctuations misidentified as signals is well below the intrinsic dark rate of the \SiPMs\ and the detector can operate as a single photon counter (see Section~\ref{sec:results}). In addition, the high bandwidth front-end electronics maintain the fast peak of the \SiPM\ signals and allow for better than \SI{20}{\nano\second} timing resolution at 1 standard deviation. Finally, the overall power consumption of the front-end electronics is less than \SI{250}{\milli\watt}, which avoids excessive heat dissipation into the detector's cryogenic operating environment. This is comparable to the cryogenic PMTs used in the DarkSide-50 detector which dissipate around \SI{100}{\milli\watt} plus an additional \SI{90}{\milli\watt} from the cryogenic pre-amplifier~\cite{Agnes:2017wf}.


\subsection{Detector overview}
The photodetector consists of 24 \SI[product-units=single]{10 x 10}{\square\milli\meter} SiPMs mounted on a substrate, referred to hereafter as a \SiPM\ tile, mated to a board housing the front-end electronics. The \SI{24}{\square\cm} \SiPM\ tile is subdivided into \SI{6}{\square\cm} quadrants and each quadrant is read out with an independent low-noise transimpedance amplifier (\TIA) based on an LMH6629 high-speed operational amplifier~\cite{TexasInstruments:2016wl}. The four \TIA\ outputs are then summed by a second stage amplifier into a single analog output that is extracted from the cryogenic environment. The \SiPM\ tile partitioning is driven by a compromise between a limit on the total power dissipation of the detector, equivalent to the total number of \TIAs, and the bandwidth and SNR attainable for a given detector capacitance. 



\subsection{Cryogenic testing setup}
The development and testing of the front-end electronics was done in a liquid nitrogen filled dewar with a flanged top. The dewar is sealed with a cover plate instrumented with a set of electrical and optical vacuum feedthroughs for power and signal lines and a calibration light source. A Keithley 2450 source measure unit was used for the \SiPM\ bias voltage and a low-noise \LNGSCryoSetupAmplifierBiasModel\ power supply was used for the front-end amplifiers.


\section{\SiPM\ Tile}
\label{sec:sipm}
\begin{figure}[!t]
  \centering
  \includegraphics[width=0.6\columnwidth]{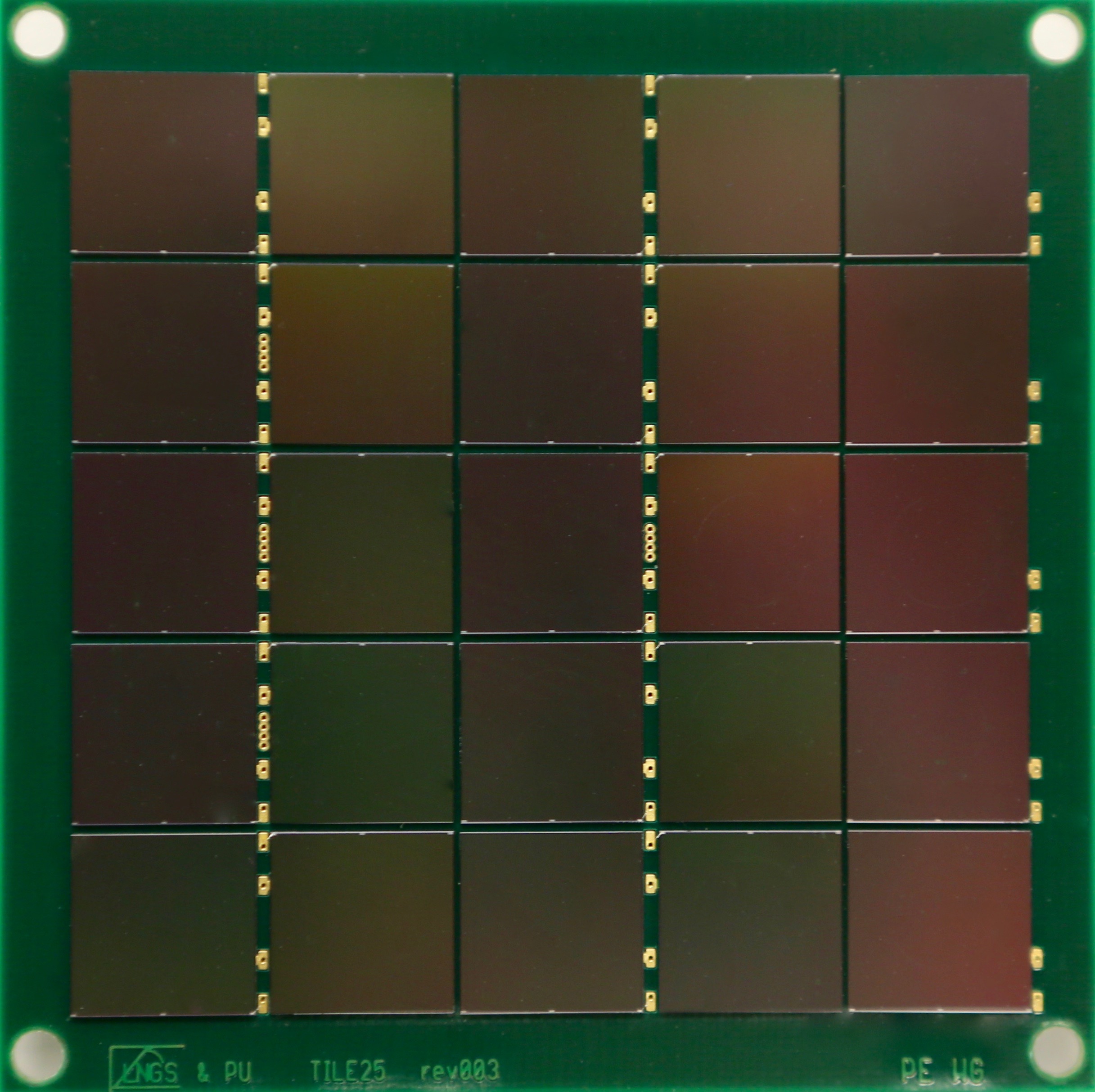}
  \caption{Picture of the test \SiPM\ tile described in the text (one \SiPM\ is disconnected).  For deployment in real detectors, the tile PCB would be re-designed to minimize any dead space, thereby increasing the overall photon detection efficiency.}
  \label{fig:tile}
\end{figure}

The \SiPM\ tile is populated with \SI[product-units=single]{10 x 10}{\square\milli\meter} NUV-HD-LF \SiPMs\ produced by Fondazione Bruno Kessler (FBK) for the DarkSide-20k collaboration~\cite{Aalseth:2017tr}. NUV-HD \SiPMs\ have a peak efficiency of about \SI{40}{\percent} in the near-ultraviolet, between \SIrange[range-phrase = --, range-units = single]{400}{420}{\nm}, and are fabricated with a cell border structure that allows for high-density packing of small cell sizes~\cite{Piemonte:2016cj}. The low-field (LF) variant of NUV-HD SiPMs have a lower field in the avalanche region, reducing the field-enhanced dark count rate relative to standard field NUV-HD \SiPMs~\cite{Ferri:2016ky}. The NUV-HD-LF devices used to fabricate the tile characterized in Section~\ref{sec:results} have \SI{25}{\micro\meter} cells and \SI{2.2}{\mega\ohm} quenching resistors at room temperature. A comprehensive study of the performance of NUV-HD-LF \SiPMs\ at cryogenic temperature can be found in~\cite{Acerbi:2017gy}: at \SI{77}{\kelvin} and with \SI{5}{\volt} of over-voltage, the darkrate is about \SI{10E-3}{cps\per\square\mm}.

The \SiPM\ parameters relevant to the design of the readout electronics are discussed in~\cite{cryo-pre} and include the shape of the signal and the equivalent capacitance ($C_d$) and resistance ($R_d$) of the device. When operated in liquid nitrogen, $C_d = \SI{4.2\pm0.1}{nF}$ and $R_d = \SI{61\pm1}{\ohm}$ for frequencies around \SI{1}{MHz}. The value of $R_d$ is negligible for frequencies above $\frac{1}{2\pi R_d C_d / 10}$ as described in~\cite{cryo-pre}.

\SiPM\ dies are mounted on an FR4 printed circuit board (PCB) using EPO-TEK EJ2189-LV conductive, silver-loaded epoxy, which has been tested for mechanical and electrical stability after repeated thermal cycling in liquid nitrogen. The epoxy is screened onto the tile in a \SI{60}{\micro\meter} thick layer, creating a coplanar surface on which the SiPM dice are placed with a manual die bonder. The top side pad of the \SiPM\ is wedge bonded with aluminum wire to a trace on the PCB. The anode and cathode of each \SiPM\ are routed to a high density Samtec LSS-150-01-L-DV-A connector on the back-side of the PCB that mates with the front-end board. 

The \SiPM\ tile is shown in Figure~\ref{fig:tile}. Neglecting the tile border, which was used to ease handling during testing, the tile has a fill factor in excess of \SI{90}{\percent}, leading to an overall photon detection efficiency (PDE) at room temperature of about \SI{35}{\percent}~\cite{Ferri:2016ky}. This can be compared to a traditional \SI{3}{\inch} cryogenic Hamamatsu R11065 PMT, which has a quantum efficiency of \SI{25}{\percent} at room temperature and an active to total surface area ratio of \SI{70}{\percent}~\cite{R11065} (and the honeycomb packing loses an other \SI{10}{\percent}).

\subsection{\SiPM\ topology}
\begin{figure}[!t]
  \centering
  \includegraphics[width=0.9\columnwidth]{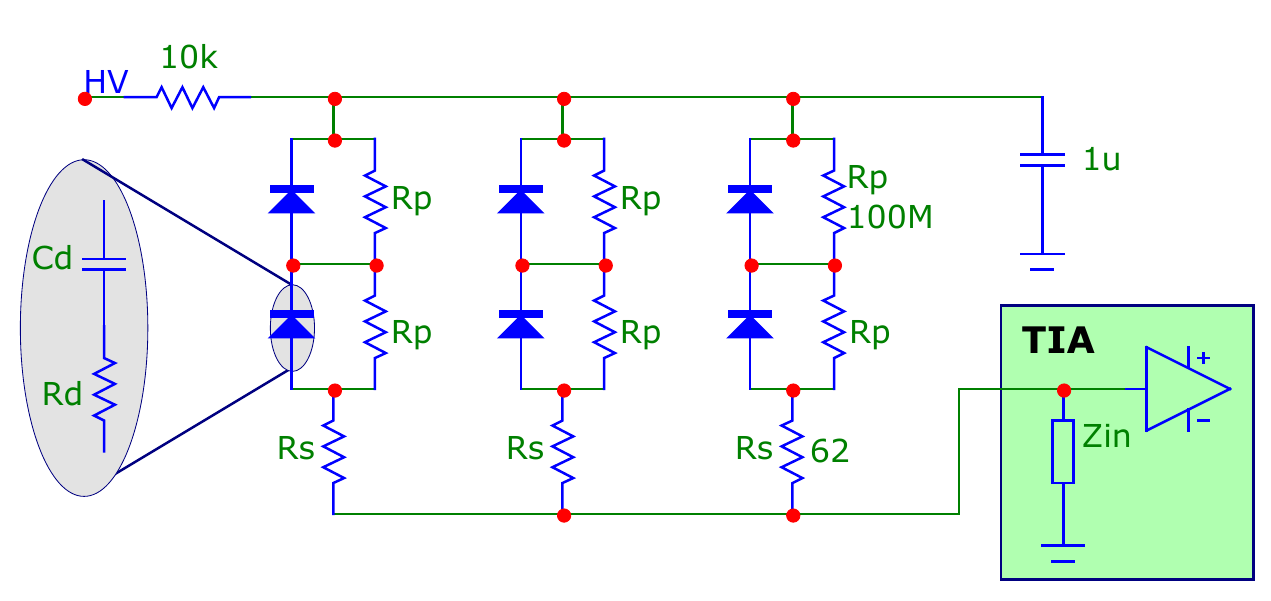}
  \caption{Schematic of the \SiPM\ layout for the \SI{6}{\square\cm} tile quadrant.}
  \label{fig:ganging}
\end{figure}

In an ideal TIA, the overall noise is proportional to the square root of the detector capacitance, $\sqrt{C_d}$\footnote{This approximation is valid when the main contribution to the input equivalent noise is the voltage noise of the amplifier ($e_n$) and the other contributions (the current noise of the amplider, $i_n$, and the Johnson-Nyquist noise of $R_f$) are negligible~\cite{sboa060}.}. In order to reduce the noise, the detector capacitance seen by the TIA can be limited by arranging the \SiPMs\ in series. However, due to capacitive coupling between the SiPMs, the signal will also be attenuated by a factor equal to the number of \SiPMs\ in series.  Theoretically, these two effects balance one another and the \SiPM\ configuration does not affect the SNR. 

In practice, several effects come into play that change the theoretical balance in favour of the configuration shown in Figure~\ref{fig:ganging} (3 branches with 2 \SiPMs\ in series each). First, the input impedance of the \TIA\ ($Z_i$) is non-zero. Therefore, a series resistance $R_s$ limits losses toward the inactive branches of the quadrant (the branches with no detected light): the relative losses can be calculated equal to $\frac{Z_i}{Z_d + R_s/2}$, where $Z_d$ is the complex impedance of the \SiPM\ (which can be formulated in terms of $R_d$ and $C_d$). This is particularly important for high frequencies, e.g., the initial discharge of the SiPM, where $Z_i$ can be large and $Z_d$ small. Second, $R_s$ and $R_d$ effectively limit the noise gain to $\frac{R_f}{(R_s + 2\,R_d)/3} + 1$, breaking the overall $\sqrt{C_d}$ noise dependence. Finally, the voltage noise at the \TIA\ input includes the Johnson-Nyquist noise of $R_d$. Increasing the number of SiPMs in series reduces the signal and increases the noise, affecting the SNR. 

The tile is subdivided into four quadrants of six \SiPMs, each readout by an independent TIA with the configuration shown in Figure~\ref{fig:ganging}. This configuration optimizes the SNR and bandwidth when compared to alternative arrangements with the same surface area.

\begin{figure}[!t]
  \centering
  \includegraphics[width=0.7\columnwidth]{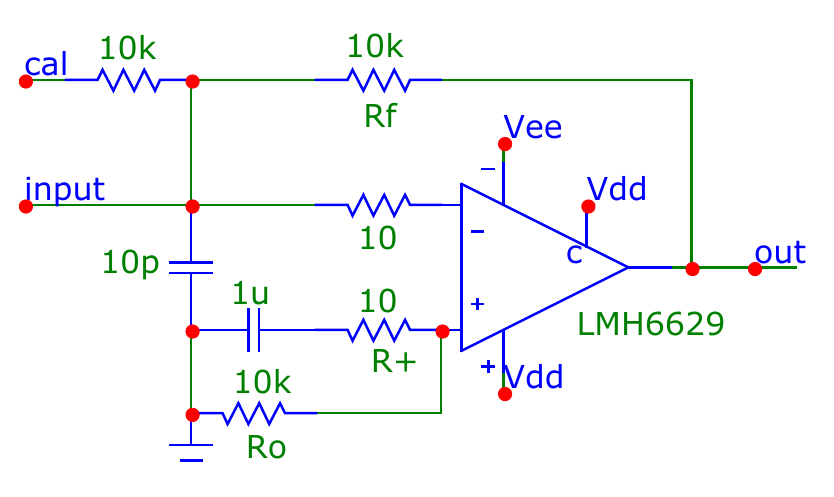}
  \caption{Schematic of the trans-impedance amplifier used for the readout of the \SI{6}{\square\cm} tile quadrant.}
  \label{fig:tia}
\end{figure}

\subsection{Bias voltage divider}
\label{subsec:divider}
At room temperature, the bias voltage across a series chain of SiPMs automatically equalizes due to the shared dark current between the devices~\cite{Ootani2013146}. The situation changes at cryogenic temperatures, where the SiPM dark rate drops below \SI{0.1}{Hz\per\square\mm}. At a gain of about \num{1d6}, the dark current is only a few pA per device. In this regime, leakage currents dominate. The quiescent current flowing through the \DSkSiPMAreaMax\ \SiPMs\ was measured to be on the order of \mbox{\SIrange[range-phrase = --, range-units = single]{0.2}{1.5}{\nano\ampere}}. 

To ensure even bias distribution between the series \SiPMs, a resistor-based voltage divider is added to the circuit as shown in Figure~\ref{fig:ganging}. Three points are critical to the design of the divider:
\begin{compactitem}
\item The precision of the divider has to compare with the desired gain uniformity ($G_U$) of the \SiPMs. For $G_U\ge\SI{95}{\percent}$ and an over-voltage to bias ratio of about $\frac{5}{32}$, the voltage partitioning accuracy has to be better than \SI{0.8}{\percent}, which requires resistors with \SI{0.5}{\percent} tolerance.
\item The current flowing through each voltage divider ($i_d$) has to be high enough so that a leakage current ($i_l$) from the \SiPMs\ will not affect the voltage partitioning  by more that the required gain uniformity. This is true when $i_d > i_l \frac{G_U}{2(1-G_U)}$. Assuming an upper limit on the leakage current at liquid nitrogen temperature of \SI{20}{\nano\ampere}, $i_d \ge \SI{200}{\nano\ampere}$.
\item The shot noise of the total divider current ($3 i_d$) can significantly contribute to the intrinsic noise of the amplifier. As described in~\cite{cryo-pre}, the current noise ($i_n$) of the LMH6629 at \SI{77}{\kelvin} corresponds to the shot noise produced by a current of about \SI{4}{\micro\ampere}. Therefore it is advisable to maintain $i_d \le \SI{1}{\micro\ampere}$.
\end{compactitem}
The \SiPM\ tile uses \SI{1}{\percent} tolerance, \SI{100}{\Mohm} resistors that were binned by their resistance at room temperature to improve the resistance variation within each divider to better than \SI{0.5}{\percent}.


\section{\SI{6}{\square\cm} electronics}
\label{sec:6cm:readout}
Each \SI{6}{\square\cm} quadrant of the SiPM tile is readout by the TIA shown in Figure~\ref{fig:tia}. This circuit is identical to the amplifier introduced in~\cite{cryo-pre} with one relevant modification. The input bias current offset of the LMH6629 at cryogenic temperature depends strongly on the power supply voltage, see~\cite{cryo-pre}, which can cause low frequency noise and instability of the output offset. The standard solution to this problem is to match $R_+$ with $R_f$ and add a bypass capacitor to filter the noise introduced by $R_+$. However, this would reduce the effectiveness of $R_+$ at dumping oscillations. Therefore, $R_+$ is left untouched and an offset compensation resistor $R_o$ equal in value to $R_f$ is added along with a filter capacitor.

\begin{figure}[!t]
  \centering
  \includegraphics[width=0.9\columnwidth]{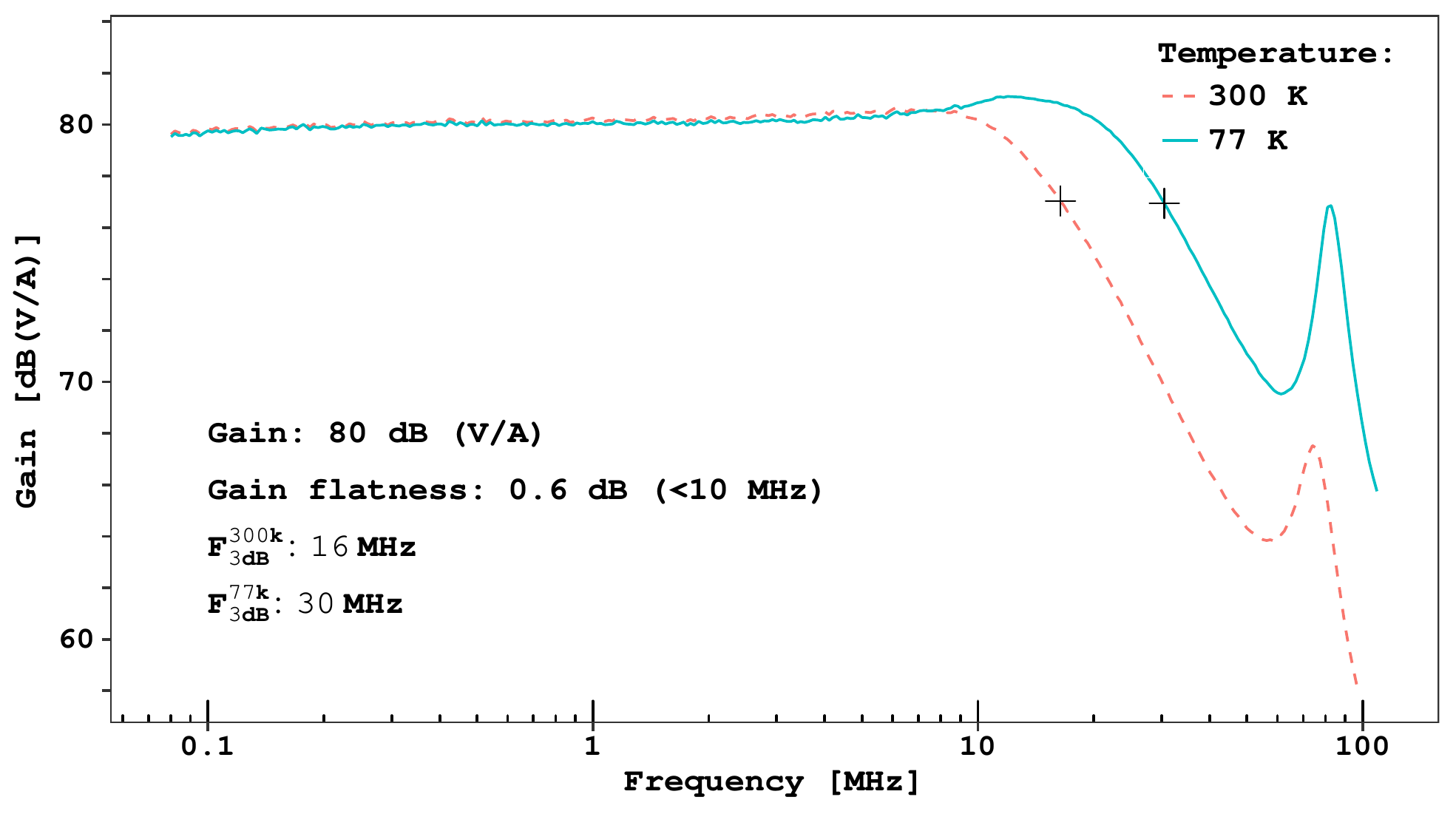}
  \caption{Bode plot from an $S_{21}$ scan of the \TIA\ coupled to a \SI{6}{\square\cm} SiPM tile quadrant at both room and liquid nitrogen temperature.}
  \label{fig:tia:bw}
\end{figure}

The gain and bandwidth of the \TIA\ coupled with a \SI{6}{\square\cm} SiPM tile quadrant are shown in Figure~\ref{fig:tia:bw}. The gain of the circuit is \SI{80}{dB(V/A)}, as expected with $R_f=\SI{10}{\kohm}$, and a flatness of \SI{0.6}{dB} up to \SI{10}{MHz}. The bandwidth in liquid nitrogen is \SI{30}{MHz}.

\subsection{Noise model}
The intrinsic voltage noise ($e_n$) of the LMH6629 at \SI{77}{\kelvin} is $\SI{0.3}{\nV/\sqrt{Hz}}$, see~\cite{cryo-pre}.  The Johnson-Nyquist voltage noise from $R_+$, $\frac{1}{3}\,R_s$ and $\frac{2}{3}\,R_d$ contribute an additional \SI{\sim0.5}{\nV/\sqrt{Hz}}, resulting in a total input voltage noise term $\SI{\sim0.64}{\nV/\sqrt{Hz}}$ (equivalent to \SI{-171}{dBm}). The voltage noise is amplified by the noise gain, which behaves like a high pass filter with $f_\textrm{3dB} \simeq \frac{1}{\pi\, (R_s + 2 R_d)\, C_d} \simeq \SI{260}{kHz}$ and an asymptotic gain of $\frac{R_f}{(R_s + 2\,R_d)/3} + 1\simeq\SI{165}{V/V}$ (\SI{\sim44}{dB}). This should result in an output noise spectrum that plateaus above \SI{1}{MHz} at \SI{-127}{dBm}.

At \SI{77}{\kelvin}, the intrinsic current noise of the LMH6629 is $i_n = \SI{1.3}{\pA/\sqrt{Hz}}$ and the shot noise of the bias voltage divider, $3\,i_d$, is negligible. This results in an output noise density of \SI{13}{\nV/\sqrt{Hz}}, corresponding to a flat spectrum at \SI{-145}{dBm}. The Johnson-Nyquist current noise of $R_f$ in liquid nitrogen results in a second flat spectrum at \SI{-151}{dBm}. The sum of these two spectra in quadrature is \SI{-144}{dBm}. This current noise component only becomes significant below \SI{10}{kHz}, where the output voltage noise drops by a few decades and, to first approximation, can be ignored.

\begin{figure}[!t]
  \centering
  \includegraphics[width=0.9\columnwidth]{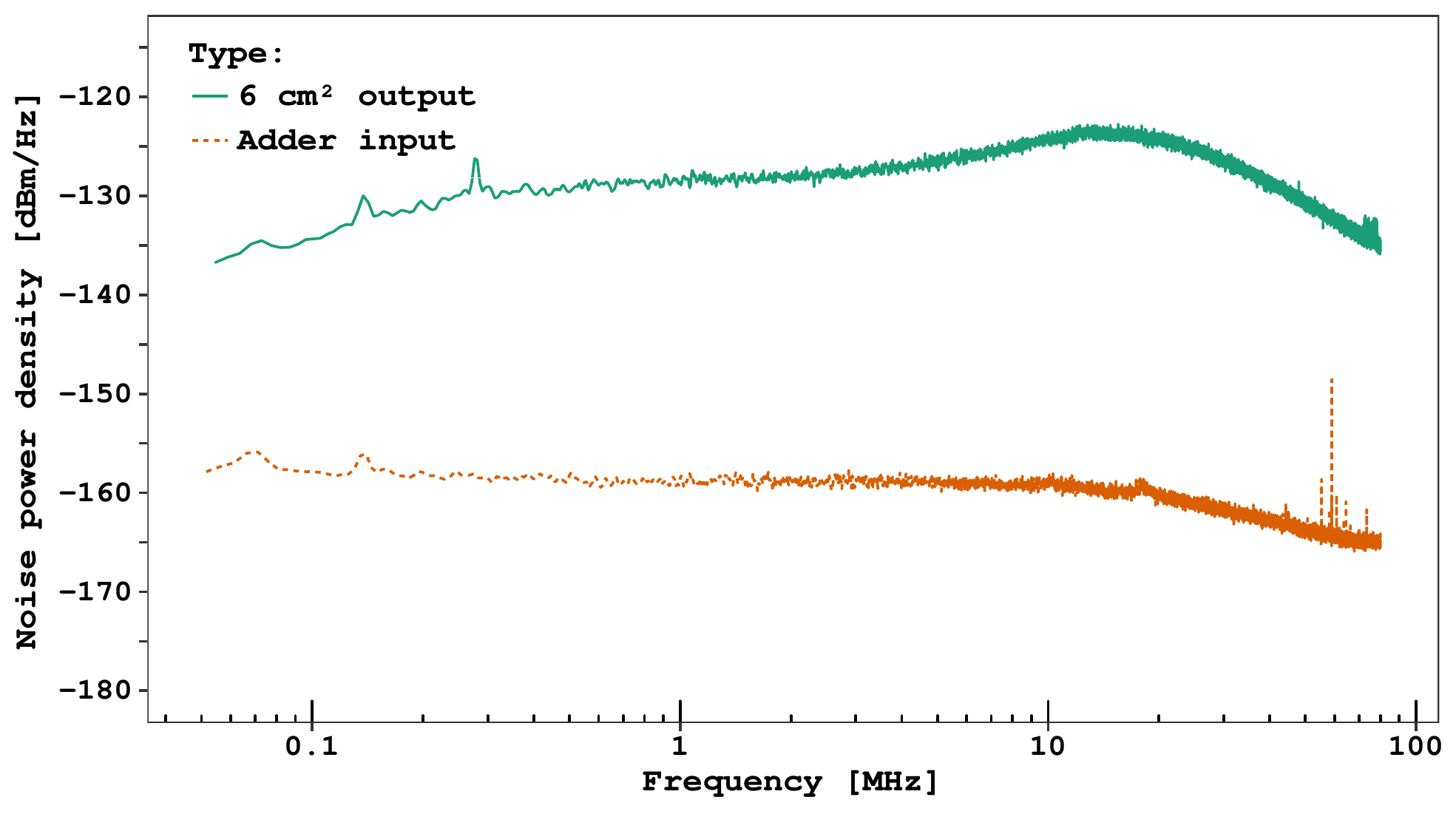}
  \caption{Output noise spectral density of the \SI{6}{\square\cm} quadrant described in the text and the input equivalent noise of the summing amplifier shown in Figure~\ref{fig:adder:schem}.}
  \label{fig:tia:noise}
\end{figure}

Figure~\ref{fig:tia:noise} shows the output noise spectrum of the \TIA\ and \SI{6}{\square\cm} SiPM tile quadrant measured with an R\&S FSV-7 spectrum analyzer at \SI{77}{\kelvin}. The features predicted by the model are present. The noise density at \SI{1}{MHz} is \SI{-128}{dBm} and $f_\textrm{3dB} \simeq \SI{200}{kHz}$. The peak at about \SI{15}{MHz} in Figure~\ref{fig:tia:noise} is not predicted by this simple noise model. The explanation of this peak is provided by the reduction of $R_d$ at high frequencies as discussed in Section~\ref{sec:sipm}.

\subsection{Performance of the \SI{6}{\square\cm} quadrant}
\label{sec:6cm:analysis}
The signal from the \TIA\, once extracted from the dewar, is further processed at room temperature by a low-noise amplifier and then digitized by a 10 bit, 1 GS/s \LNGSCryoSetupDigitizerModel\ configured with a \SI{5}{\micro\s} pre-trigger and a \SI{15}{\micro\s} total gate time. The digitizer was triggered in coincidence with a light pulse from a Hamamatsu PLP-10 sub-nanosecond \SI{405}{\nano\meter} laser source. The output of the PLP-10 was attenuated so that an average of one photon was incident on the detector per pulse. The \SiPMs\ were operated at an over-voltage of \SI{5}{V}, corresponding to a gain of  \SI{1.5E6}{\coulomb\per\coulomb}.

\begin{figure}[!t]
  \centering
  \includegraphics[width=0.95\columnwidth]{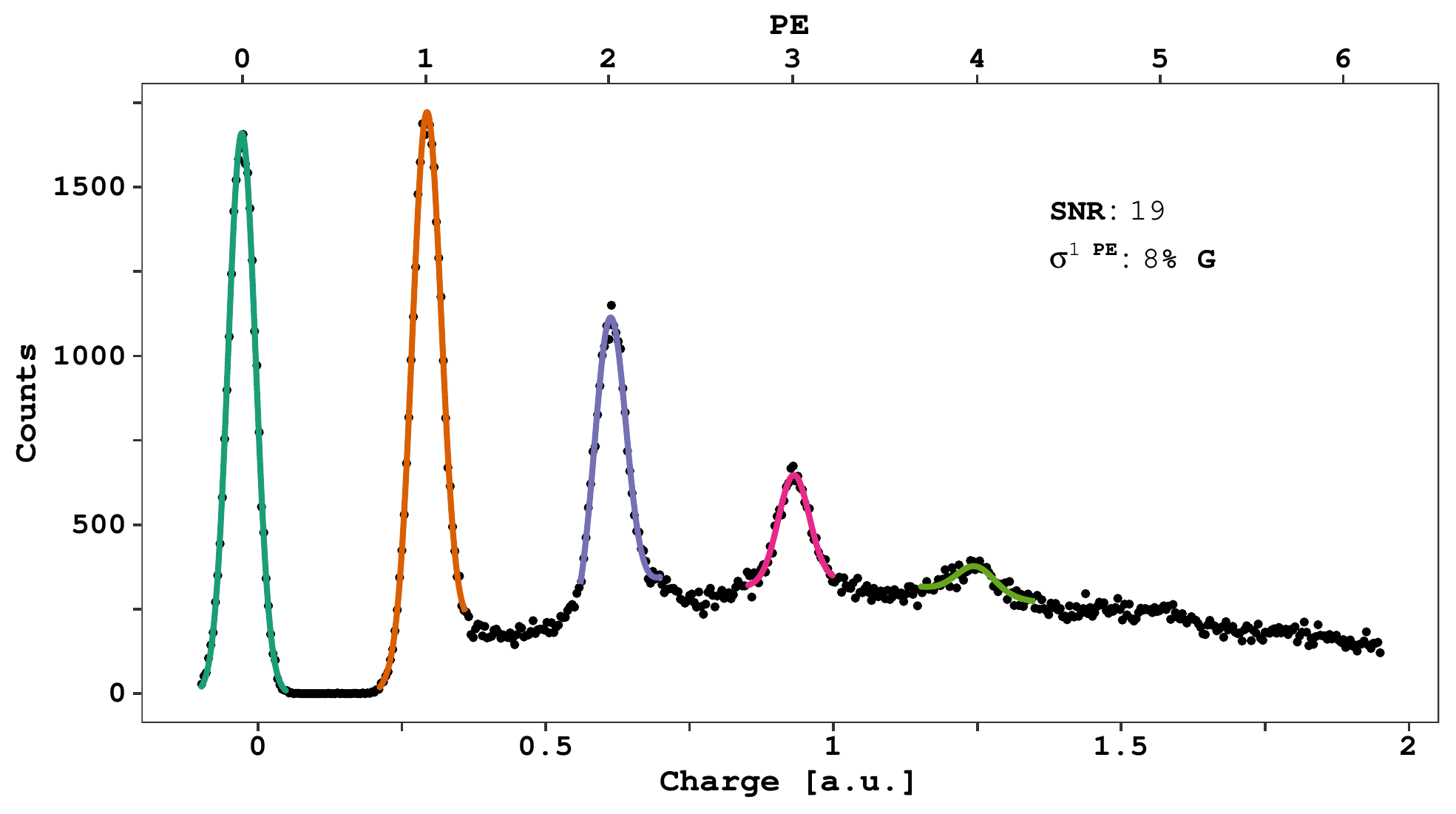}
  \caption{Photoelectron spectrum from a \SI{6}{\square\cm} detector quadrant calculated using a fixed window integration. The solid lines represent gaussian fits to the photoelectron and baseline peaks.}
  \label{fig:snr:6cm2:int}
  \vskip 3mm
  \includegraphics[width=0.95\columnwidth]{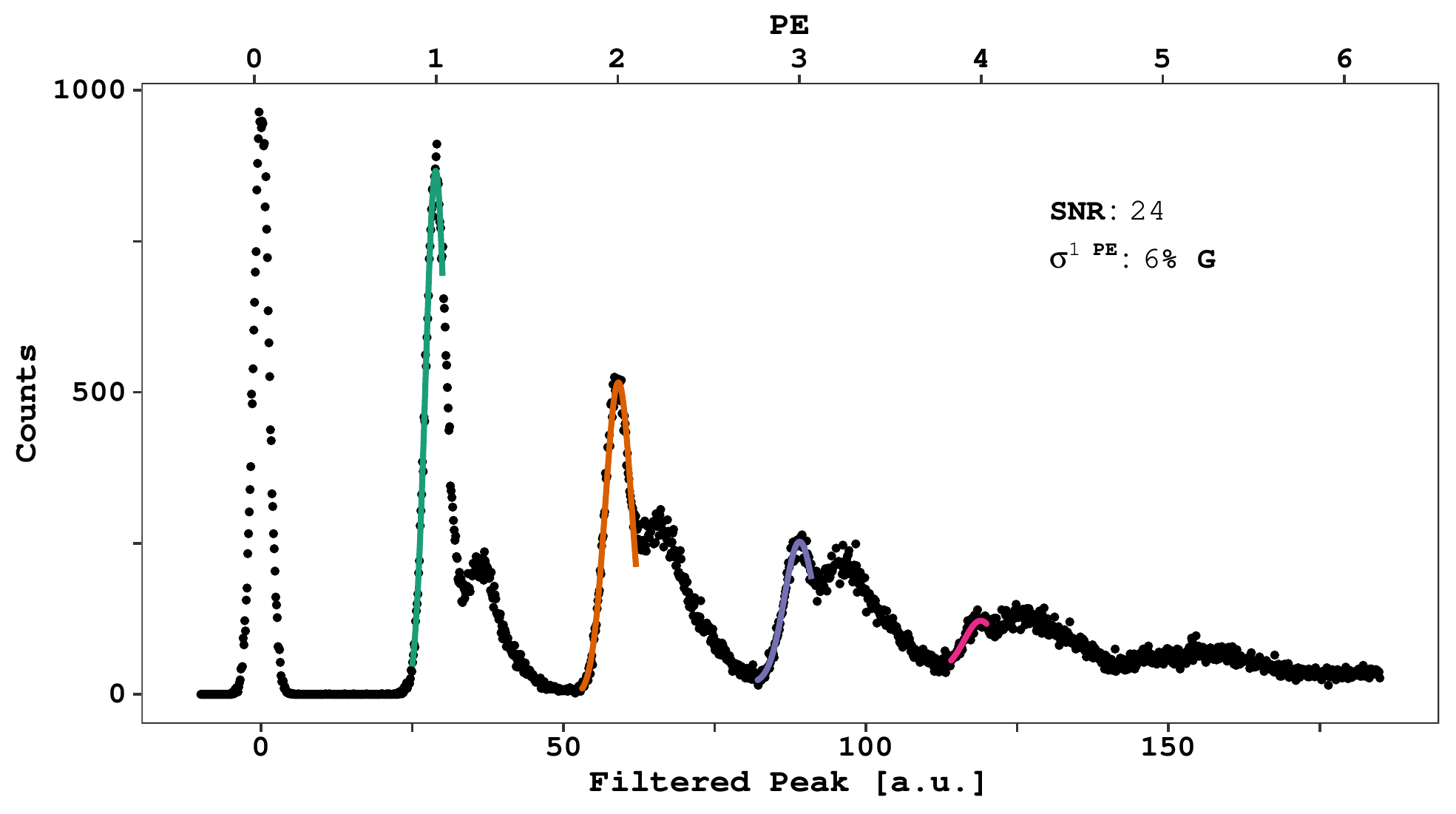}
  \caption{Photoelectron spectrum of a \SI{6}{\square\cm} detector quadrant calculated with a matched filter. The solid lines represent a gaussian fit to the low energy portion of the photoelectron peaks. The high energy tails of the peaks are distorted by events that contain an after-pulse.}
  \label{fig:snr:6cm2:fil}
\end{figure}

The amplitudes of the digitized waveforms were calculated offline using two different methods. Figure~\ref{fig:snr:6cm2:int} shows the spectrum obtained using a fixed integration window of \SI{2.2}{\micro\second} (\SI{4}{\tau}). The baseline noise is \SI{5}{\percent} of the gain and the separation between the photoelectron peaks is limited by after-pulsing of the \SiPMs. Figure~\ref{fig:snr:6cm2:fil} shows the spectrum obtained using a matched filter algorithm~\cite{Turin:1960kw}, which results in an SNR of \num{24}. The matched filter more clearly separates events with after-pulses from the photoelectron peaks, resulting in non-gaussian peak shapes.

The gain uniformity of the first photoelectron peak in Figure~\ref{fig:snr:6cm2:fil} can be calculated by subtracting in quadrature the baseline noise from $\sigma_{1\textrm{PE}}$, resulting in $G_U = \SI{96}{\percent}$, comparable to the specification outlined in Section~\ref{subsec:divider}

\section{Cryogenic Summing Amplifier}
\label{sec:adder}
The output signals of the four \TIAs\ from the four \SI{6}{\square\cm} quadrants are combined at the input of a cryogenic summing amplifier that must amplify their sum for transmission across the dewar while maintaining the signal bandwidth, noise and dynamic range of the \SI{6}{\square\cm} readout.


\subsection{LMH6624 Cryogenic Characterization}
\begin{figure}[!t]
  \centering
  \includegraphics[width=0.9\columnwidth]{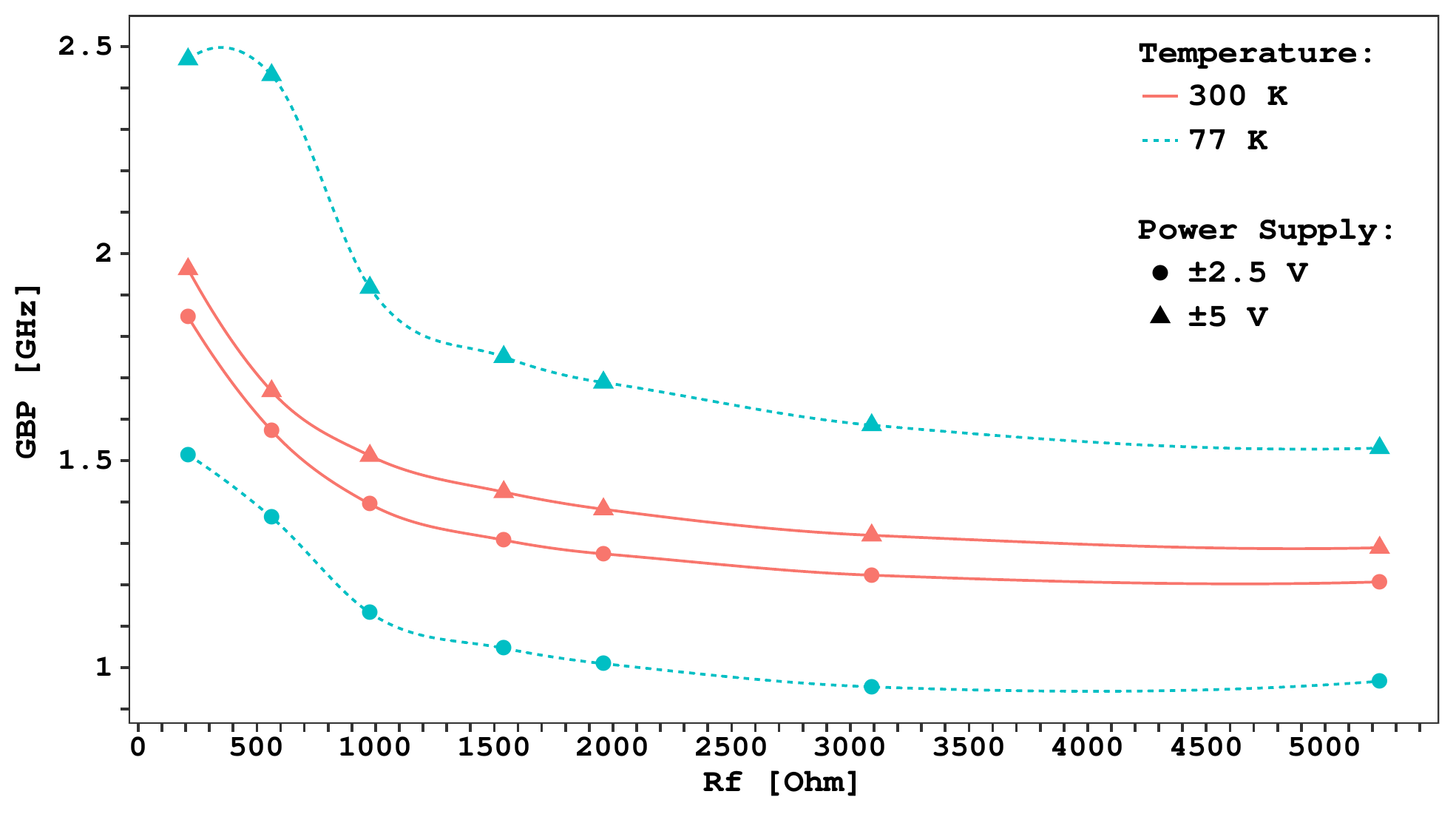}
  \caption{Gain bandwidth product of the LMH6624 test circuit in the non-inverting configuration versus feedback resistor value at room temperature and in liquid nitrogen ($\textrm{gain} = R_f/\SI{25}{\ohm} + 1$). The lines are drawn to guide the eye.}
  \label{fig:gbp}
\end{figure}
The core of the summing amplifier is an LMH6624 operational amplifier from Texas Instruments.  At room temperature, the LMH6624 has a gain bandwidth product of \SI{1.5}{GHz} and an input voltage noise of \SI{0.92}{\nV/\sqrt{Hz}}~\cite{TexasInstruments:6624}. The LMH6624 was characterized in a liquid nitrogen bath following the procedure described in~\cite{cryo-pre}. The input voltage noise density at \SI{77}{\kelvin} drops to \SI{0.53\pm0.02}{\nV/\sqrt{Hz}}. Figure~\ref{fig:gbp} shows the gain bandwidth product of the test circuit versus $R_f$. At room temperature, the change in the gain bandwidth product due to the power supply voltage is relatively small, while in liquid nitrogen the variation is almost a factor of two. The overall shape of the curves is due to stray capacitance in the feedback path. The \SI{1}{\decibel} output compression point measured in liquid nitrogen is \SIrange[range-phrase = ~and~, range-units = single]{2.2}{5.8}{\volt_\textrm{PP}} (before back-termination) with a quiescent current of \SIrange[range-phrase = ~and~, range-units = single]{3}{5}{mA} respectively for \SIrange[range-phrase = ~and~, range-units = single]{\pm2.5}{\pm5}{\volt} power supplies.

\subsection{Summing amplifier design}
\begin{figure}[!t]
  \centering
  \includegraphics[width=0.8\columnwidth]{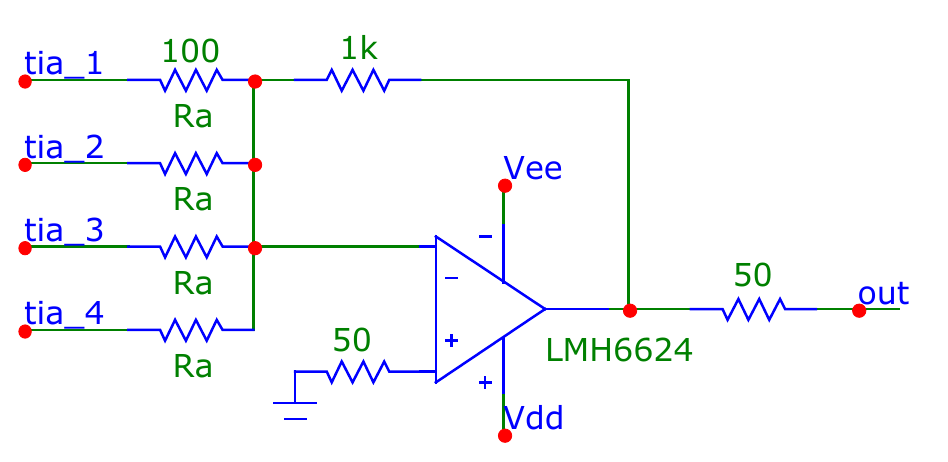}
  \caption{Schematic of the summing amplifier used to combine the outputs of the SiPM tile quadrants.}
  \label{fig:adder:schem}
\end{figure}

The summing circuit is shown in Figure~\ref{fig:adder:schem}. The input resistors ($R_a$) are \SI{100}{\ohm} to avoid over-loading the LMH6629 output stage. The signal gain is \SI{10}{V/V} and the noise gain is \num{41}~\cite{MT045}. The additional factor of ten amplification boosts the single photoelectron signal into the tens of millivolt range, simplifying the signal extraction and acquisition. The operational amplifier uses the same \SI{\pm2.5}{V} power supply as the \TIAs.

The bandwidth of the circuit was measured with a Vector Network Analyzer (VNA) configured for S21 scan to be \SI{36}{MHz} at room temperature and \SI{30}{MHz} in liquid nitrogen. The electronic noise of the summing amplifier is dominated by the intrinsic voltage noise of the LMH6624 multiplied by the {\SI{41}{V/V}} noise gain and the Johnson-Nyquist noise of the input resistors. The overall input equivalent noise of the amplifier is $\frac{41}{10} \sqrt{\left(e_n^2 + 4 k_B T \frac{R_a}{4}\right) f_{3dB} \frac{\pi}{2}}  \simeq \SI{17}{\micro\volt}$. Figure~\ref{fig:tia:noise} compares the measured input equivalent noise spectrum of the amplifier to the output noise of the \SI{6}{\square\cm} quadrant readout, demonstrating that the summing amplifier provides effectively no contribution to the overall noise of the system. In addition, the summing amplifier is slightly faster than the \TIA\ and does not affect the bandwidth of the circuit. A \SI{50}{\ohm} back-termination is added to avoid signal reflections on the coaxial cable: its net effect is a factor of two reduction in gain and electronic noise.

\subsection{Offset and dynamic range}
The coherent sum of the current through all of the bias voltage dividers can produce a considerable offset at the output of the summing amplifier, which can be evaluated as $12\, i_d\times R_f \times 10/2$. In the present configuration, this becomes \SI{+0.2}{\volt} after back-termination. Unfortunately, the amplified photoelectron signal is also positive (after two inversions) and the offset limits the dynamic range of the system to a few tens of photoelectrons. This was not a limitation for the present goal of performing low noise readout of a few photoelectrons. However, because offset contribution of the bias voltage divider is well known in advance, it is possible to bias the non-inverting pin of the amplifier and cancel the offset (or even further increase the negative offset). As an alternative, the authors are studying a rail to rail differential solution that would maximize the dynamic range. 

\begin{figure}[!t]
  \centering
  \includegraphics[width=0.95\columnwidth]{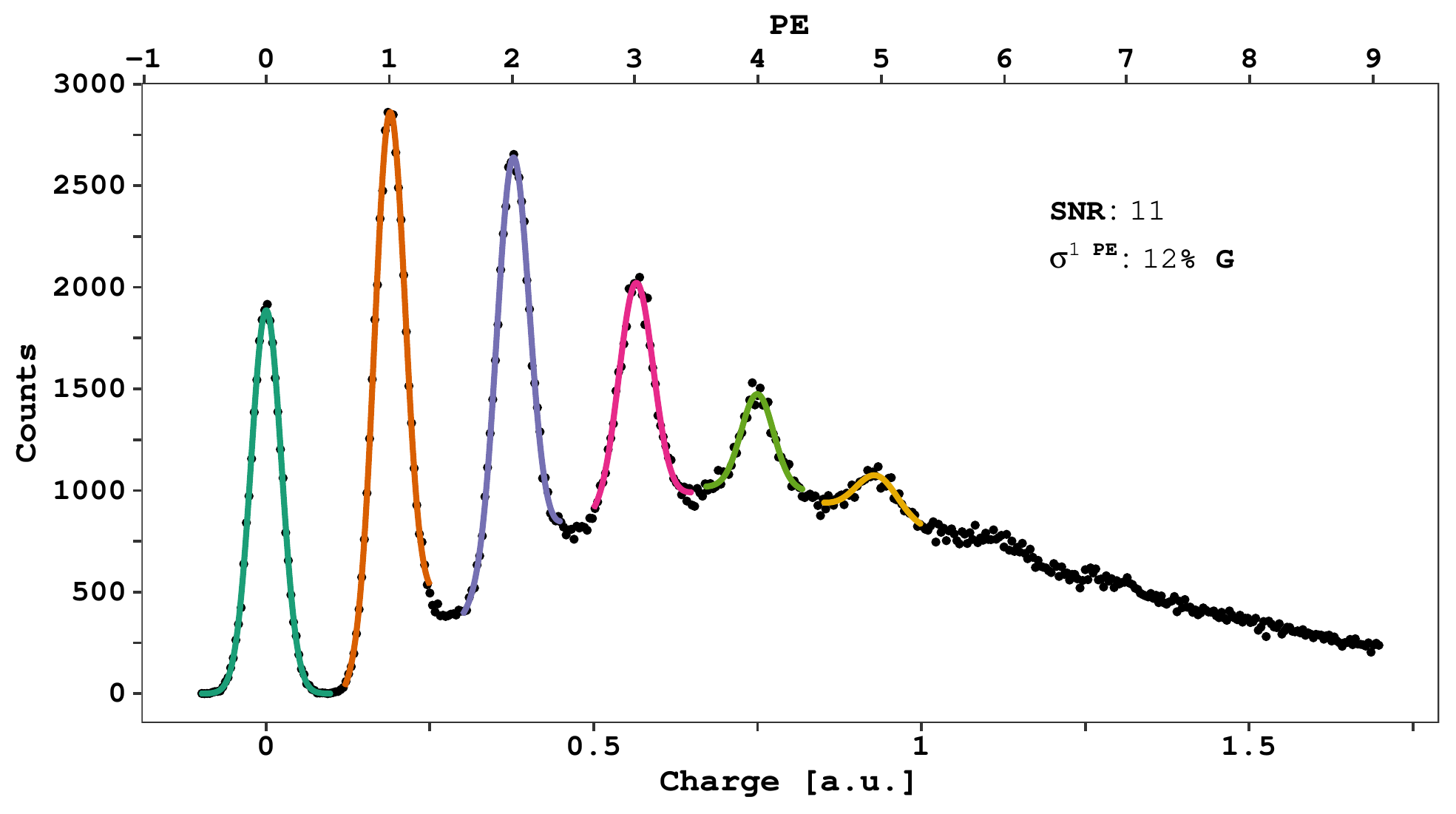}
  \caption{Photoelectron spectrum of the full \SI{24}{\square\cm} detector calculated using a fixed window integration. The solid lines represent a gaussian fit to the photoelectron and baseline peaks.}
  \label{fig:snr:24cm2:int}
  \vskip 3mm
  \includegraphics[width=0.95\columnwidth]{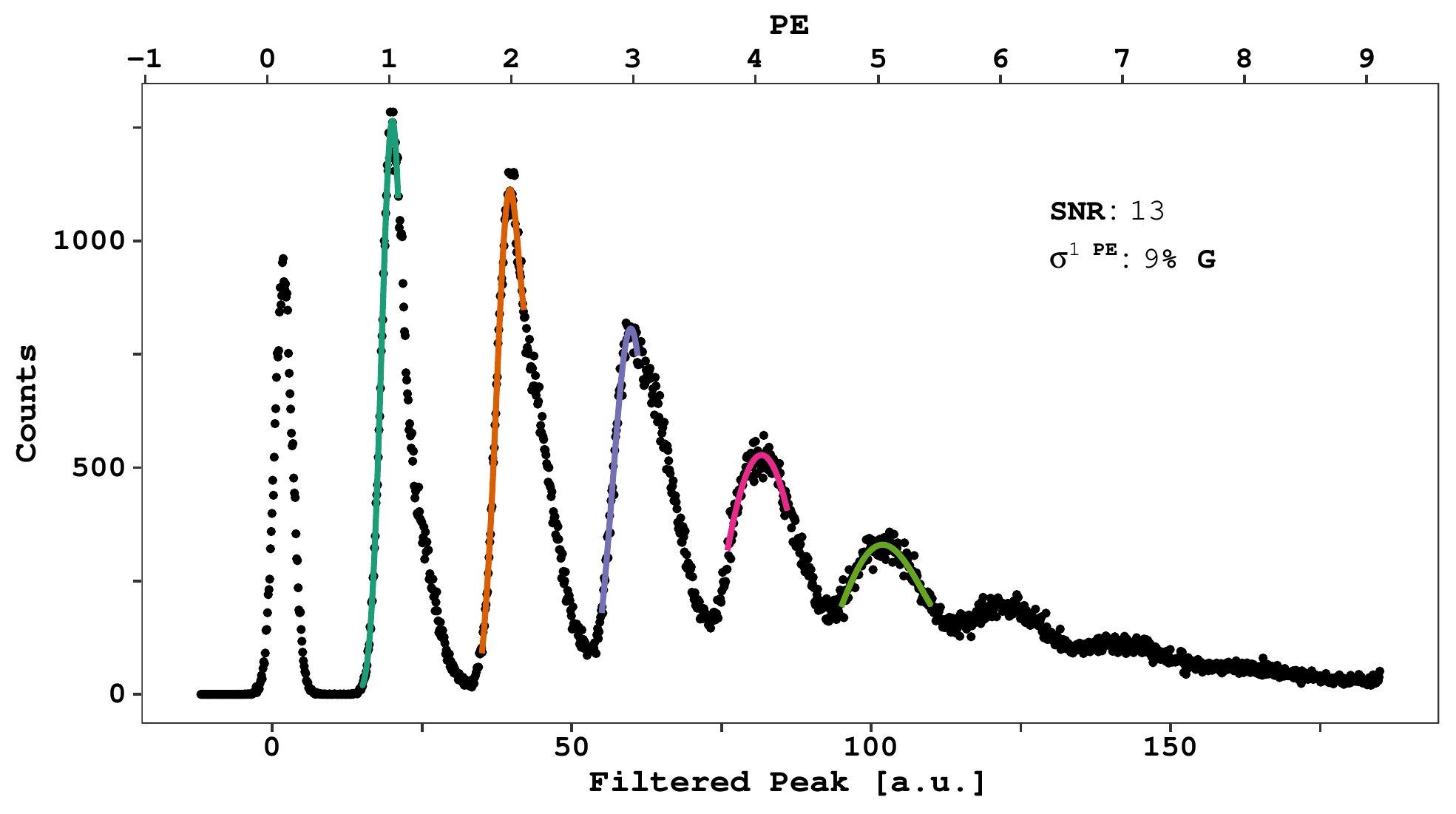}
  \caption{Photoelectron spectrum of the full \SI{24}{\square\cm} detector calculated using a matched filter. The solid lines represent a gaussian fit to the photoelectron peaks.}
  \label{fig:snr:24cm2:fil}
\end{figure}

\subsection{Detector performance}
\label{sec:results}

The signal extraction from the summing amplifier, its digitization and the analysis algorithms follow the methods used for the \SI{6}{\square\cm} detector described in Section~\ref{sec:6cm:analysis}. The spectrum from the fixed window integration is shown in Figure~\ref{fig:snr:24cm2:int} and the matched filter spectrum is shown in Figure~\ref{fig:snr:24cm2:fil}. In both cases, the SNR is greater than 10 and the gain uniformity meets the design goal, $G_U = \SI{95}{\percent}$.

\begin{figure}
 \centering
  \includegraphics[width=\columnwidth]{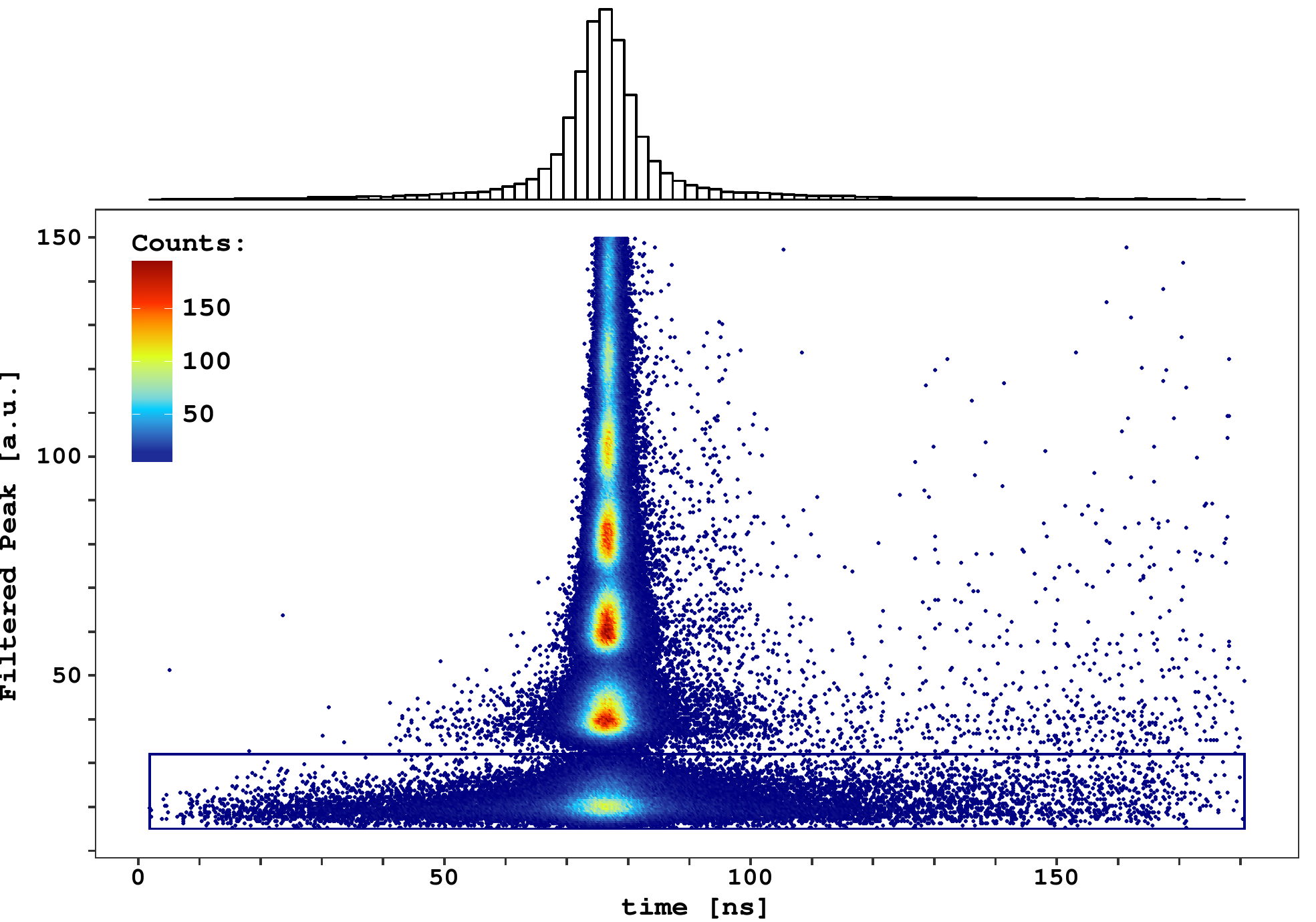}
  \caption{Peak amplitude of the matched filtered waveform versus the reconstructed time relative to the laser sync. The top histogram shows the timing distribution of the single photoelectron peak (within the black box). The single photoelectron peak timing standard deviation is \SI{16}{\nano\second}.}
  \label{fig:24amp_vs_t}
\end{figure}

The timing of an event can be extracted from the peak time of the matched filtered waveform, as described in~\cite{cryo-pre}. Figure~\ref{fig:24amp_vs_t} shows the event amplitude versus the reconstructed event time. The distribution of the reconstructed arrival time of the single photoelectron peak (top panel of Figure~\ref{fig:24amp_vs_t}) contains a narrow central peak, with a standard deviation of about \SI{4}{\nano\second} and non-gaussian tails that deteriorate the overall timing resolution to \SI{16}{\nano\second}. These tails account for less than \SI{10}{\percent} of the total events and are due to photoelectrons whose raw waveforms differ from the matched filter template. This is due to noise pickup jittering around the digitizer trigger, noise that, when coincident with the photoelectron signal, distorts the fast rising edge of the waveform and broaden the timing resolution.

\section{Conclusions}
\label{sec:conclusions}
We report on the first implementation of electronics for a large area (\SI{24}{\square\cm}) single-channel, SiPM-based cryogenic photodetector with single photon sensitivity.  This is achieved despite the large \SiPM\ capacitance by subdividing the detector into four \SI{6}{\square\cm} sub-arrays individually coupled to custom designed cryogenic transimpedance amplifiers. Signals from the four transimpedance amplifiers are then summed with a cryogenic summing amplifier to obtain a single-channel readout. Signals collected from the detector were analyzed using a matched filtering technique, resulting in an excellent signal to noise ratio (\num{> 10}) and timing resolution (\SI{< 20}{\nano\second}). Moreover, the excellent overall PDE ($\sim$\SI{35}{\percent}) and very low dark rate ($\sim$\SI{24}{cps}) make these devices ideal candidates for future, low-background cryogenic dark matter and neutrino detectors.

\bibliographystyle{IEEEtran}
\bibliography{$TEXMFHOME/ds,tile-24}
\end{document}